\documentstyle[preprint,floats,aps,epsfig]{revtex}
\begin{document}
%\baselinestretch{5}
\tightenlines

\title{Effect of Nuclear Deformation on $J/\psi$ Suppression\\
in Relativistic Nucleus-Nucleus Collisions}
\vspace{0.1in}
\author{Ben-Hao Sa$^{1,2}$\footnotemark and An Tai$^3$}

\address{
$^1$  Cyclotron Institute, Texas A$\&$M University, College Station,
      TX 77843-3366 \\
$^2$  China Institute of Atomic Energy, P. O. Box 275 (18),
Beijing, 102413 China \\
$^3$  Department of Physics and Astronomy, University of California at 
      Los Angeles, Los Angeles, CA 90025
\footnotetext{E-mail: sa@kogroup.tamu.edu; sabh@iris.ciae.ac.cn}
}

\maketitle
\begin{abstract}

Using a hadron-string cascade model, JPCIAE, we study the
effect of nuclear deformation on $J/\psi$ suppression in the
collision of uranium nuclei at 200A GeV/c. 
We find that the $J/\psi$ survival probability is much smaller if
the major axes of both deformed nuclei are along
the beam direction than if they are perpendicular to the beam
direction.

\noindent{PACS numbers:  25.75.Dw, 24.10.Lx, 24.85.+p, 25.75.Gz}
\end{abstract}
\vspace{0.1in}

Because of color screening, $J/\psi$ is expected to dissociate in the 
quark-gluon plasma (QGP) formed in relativistic heavy ion collisions 
\cite{masa}. The resulting suppression of $J/\psi$ production 
in these collisions has been one of the most studied signals for 
the formation of the quark-gluon plasma. Experiments at CERN have 
indeed shown that $J/\psi$ production is reduced in both proton-nucleus 
\cite{na3} and heavy ion collisions \cite{na381,mic1} compared to that
expected from the superposition of proton-proton collisions at 
same energies. For collisions involving light projectiles such as
p-A \cite{na3}, O-U, and S-U collisions \cite{na381}, conventional
mechanisms of $J/\psi$ absorption by nucleons \cite{cap1,huf1} and
comovers \cite{gyu1,vog1,gre1,vog2,huf2,cy1,bla1,kha1,vog3,kha2}
seem to be sufficient in accounting for the measured suppression.
On the other hand, in collisions with heavy projectiles such as
the Pb-Pb collision \cite{mic1}, whether the measured anomalously 
large $J/\psi$ suppression in central collisions can be explained 
by hadronic absorption alone is still under debate \cite{vog4,cap2,vog5}, 
and explanations based on QGP effects have also been proposed
\cite{cy1,bla1,kha2,shu1,gre2}.

To study microscopically $J/\psi$ production and absorption in
heavy ion collisions, transport models have been used
\cite{ko1,cas1,sa1,cas2,sa2,kah1,gal1,gre3}. Although there are differences  
among them, it has been commonly found that the explanation of anomalous 
J/$\psi$ suppression in Pb-Pb collisions needs to introduce new 
mechanism besides hadronic absorptions. It has been found in Refs. 
\cite{cas2,sa2} that the inclusion of dissociation of the
pre-$J/\psi$ $c\bar c$ state by the color electric field in the
initial dense matter can also explain
the anomalous $J/\psi$ suppression in Pb-Pb collisions. 
Whether a QGP is formed in heavy ion collisions at CERN SPS is still  
needed to have further studies. One suggestion is to 
study $J/\psi$ production in collisions of deformed nuclei \cite{shu2} 
as the large spatial anisotropy created even in the central collisions
of these nuclei offers the possibility to study the mechanisms for 
$J/\psi$ suppression from their final azimuthal distribution \cite{heis}.
Also, it has been shown in Ref. \cite{bao} that for 
collisions of deformed nuclei at AGS energies, the maximum energy
density reached in the initial stage is higher (about 38\%) and
also lasts longer if the major axes of both nuclei are along the
beam direction (tip-tip) than if they are perpendicular to the
beam direction (body-body). In high energy density matter,
$J/\psi$ is dissociated due to either color screening if the matter
is a quark-gluon plasma or the color electric field if it is a
string matter. In both cases, we expect that $J/\psi$ suppression
will be more appreciable in the tip-tip collision than in the
body-body collision. In this article, we shall report the results
of our study of this effect in U-U collisions at SPS energies
using a hadron-string cascade model, JPCIAE \cite{sa1,sa2},
which has been shown to give a satisfactory description of the
$J/\psi$ suppression data from collisions of spherical nuclei at
SPS energies. As we shall show below, because of increasing energy
density and collision time, a more pronounced $J/\psi$ suppression
due to the color electric dissociation is indeed seen in the
tip-tip collision than in the body-body collision.

The JPCIAE model is an extension of the LUND string model
\cite{sjo1} to include $J/\psi$ production and absorption. In this
model, a nucleus-nucleus collision is depicted as a superposition
of hadron-hadron collisions. If the center-of-mass energy
of a hadron-hadron collision is larger than certain value, e.g., $\geq$4
GeV, the PYTHIA routines are called to describe this
interaction. Otherwise, it is treated as a conventional two-body
interaction as in the usual cascade model \cite{cug1,ber1,sa3}.
Furthermore, for hadron-hadron collisions with center-of-mass 
energy above 10 GeV, a $J/\psi$ is produced using the PYTHIA 
routines through the reaction
\begin{equation}
g + g \rightarrow J/\psi + g,
\end{equation}
where $g$ denotes gluons in a hadron. Final-state interactions among 
produced particles and the participant and spectator nucleons are taken
into account by the usual cascade model.

Mechanisms for $J/\psi$ suppression in the JPCIAE model include
the hadronic absorption by both baryons and mesons, the energy
degradation of leading nucleons, and the dissociation of the
$J/\psi$ precursor state of $c\bar c$ pair in the color electric
field of strings \cite{sa2}. The total $J/\psi$ suppression factor
in JPCIAE is thus given by
\begin{equation}
S_{\rm total}^{J/\psi} = S_{\rm abs}^{J/\psi}\times
S_{\rm deg}^{J/\psi}\times S_{\rm dis}^{J/\psi},
\end{equation}
where $S_{\rm abs}$, $S_{\rm deg}$, and $S_{\rm dis}$ denote,
respectively, the suppression factor due to the above three
mechanisms. We note that the first two mechanisms have also been
employed recently in Ref. \cite{gal1} to study $J/\psi$ production
in proton-nucleus collisions at the Fermilab energies.  Details of
the JPCIAE model can be found in Refs. \cite{sa1,sa2}.

For the U-U collision, the initial distribution of the projectile
and target nucleons is assumed to be uniform inside an ellipsoid
with a major semi-axis $a=R(1+2\delta/3)=8.4$ fm and a minor
semi-axis $b=R(1-\delta/3)=6.5$ fm if we use a deformation
parameter $\delta=0.29$ and an equivalent spherical radius $R=7.0$
fm \cite{bomo}. Other parameters in the model are kept the same as
before \cite{sa1,sa2}, i.e., $\tau=1.2$ fm/c for the formation
time of produced particles, $\sigma^{\rm abs}_{J/\psi-B}=6$ mb and
$\sigma^{\rm abs}_{J/\psi-M}=3$ mb for the $J/\psi$ absorption
cross sections by baryons and mesons, respectively, and
$c_s=6.0\times 10^{-7}$ GeV$^{-1}$ for the effective color
electric dissociation coefficient which was determined from
fitting the $J/\psi$ suppression data from the Pb-Pb collision at
158A GeV/c \cite{sa2}.

\begin{figure}[ht]
\centerline{\hspace{-0.5in}
\epsfig{file=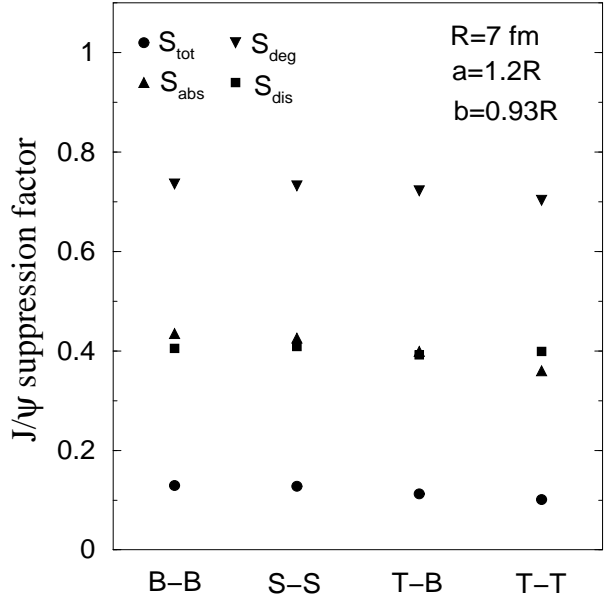,width=3.1in,height=3.1in,angle=-90}}
\vspace{0.2in}
\caption{$J/\psi$ suppression factor in central U-U
collisions at 200A GeV/c and different orientations: triangles for
the nuclear absorption $S_{\rm abs}$, inverted triangles for the
energy degradation $S_{\rm deg}$, squares for dissociation by the
color electric field $S_{\rm dis}$, and solid circles for the
total suppression factor $S_{\rm total}$.} \label{jp}
\end{figure}

In Fig. \ref{jp}, we show by solid circles the calculated total
$J/\psi$ suppression factor in central U-U collisions at 200A
GeV/c for different orientations of the colliding nuclei. The
results labeled by body-body (B-B), tip-body (T-B), and tip-tip
(T-T) correspond, respectively, to collisions in which both minor,
one major and one minor, and both major axes of the projectile and
target nuclei are parallel to the beam direction. For comparison,
we also show the results from treating both nuclei as spherical
(S-S). We see that the $J/\psi$ survival probability decreases as
the orientation changes from B-B to S-S, to T-B, and to T-T. This
result can be understood qualitatively from the dependence of the
passing time between the two colliding nuclei and the number of
participant nucleons on their orientation \cite{shu2,nu1}. For
central collisions, while the number of participant nucleons
is essentially the same for the four collision geometries, the 
nuclear passing times are, however, different.  For the B-B, S-S, 
T-B, and T-T collisions, the nuclear passing times are approximately 
given by $t^{B-B}=2b/(\beta \gamma)$,
$t^{S-S}=2R/(\beta \gamma)$, $t^{T-B}=(a+b)/(\beta \gamma)$, and
$t^{T-T}=2a/(\beta \gamma)$, respectively. In the above, $\beta$
and $\gamma$ are, respectively, the velocity and Lorentz factor in
the nucleon-nucleon center-of-mass frame. Relative to the passing
time for the collision of spherical nuclei, the following ratio is
obtained: $t^{B-B}:t^{S-S}:t^{T-B}:t^{T-T}=0.93:1.0:1.07:1.2$.
The 28\% decrease in the
$J/\psi$ suppression factor for the T-T collision compared to that
for the B-B collision, as shown in Fig. \ref{jp}, is close to the
30\% difference between the passing time in these collisions.

The effect of deformation on each of the $J/\psi$ suppression
mechanisms is also shown in Fig. \ref{jp} by triangles for $S_{\rm
abs}$, inverted triangles for $S_{\rm deg}$, and squares for
$S_{\rm dis}$. It is seen that the hadronic absorption is
increased by about 21\% in the T-T collision than in the B-B
collision. Effects due to the energy degradation of leading
nucleons only leads to a 5\% more suppression in the T-T than in
the B-B collision. This small deformation effect results from the
fact that $J/\psi$'s are mainly produced in first
nucleon-nucleon collisions. No deformation effect is seen from the
dissociation by the effective color electric field of strings. That is because 
the dissociation probability of $J/\psi$ precursor state of $c\bar c$ 
pair in initial dense string matter is assumed to be depending on the 
energy, centrality and size of collision system as 
\begin{equation}\label{dis}
P_d = c_s\sqrt{s_{\rm NN}}e^{-(b/R_L)^2}AB
\end{equation}
($S_{\rm dis}^{J/\psi}$=1-$P_d$), where $\sqrt{s_{\rm NN}}$ 
being the initial center-of-mass energy  
of two colliding nucleons, $b$ is the impact parameter of the nucleus-nucleus 
collision, and $R_L$ is the radius of the larger of the projectile and
target nuclei with atomic number $A$ and $B$, respectively \cite{sa2}. The 
above dissociation probability is proposed based on the continuous excitation 
picture of strings in LUND model. The color electric field is built up
along a string through binary nucleon-nucleon (or nucleon-string, string-
string) collisions. The more 
such collisions a string is experienced, the stronger color electric field
will be formed along the string, thus the more likely a $J/\psi$ would be
dissociated by such a color electric field. In above 
calculations we have used the same dissociation coefficient in Eq. (\ref{dis}) 
, irrespective of the energy density reached in initial string matter 
might be different among the orientations of the colliding nuclei.

\begin{figure}[ht]
\centerline{\hspace{-0.5in}
\epsfig{file=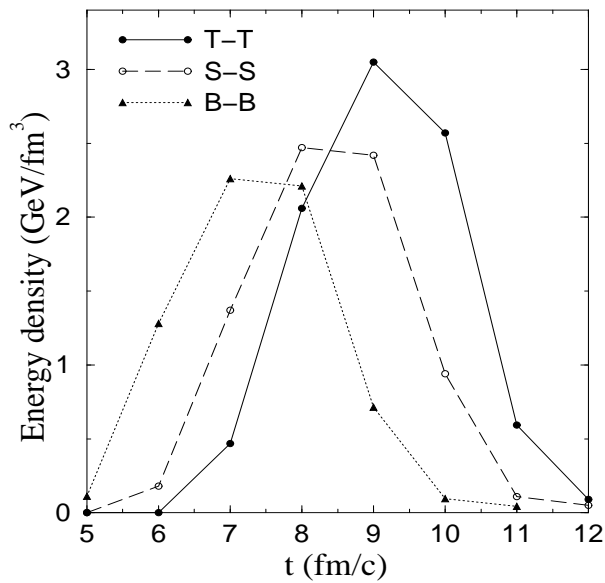,width=3.1in,height=3.1in,angle=-90}}
\vspace{0.2in}
\caption{Time evolution of central energy density
in the U-U collision at 200A GeV/c.}
\label{energy}
\end{figure}

\begin{figure}[ht]
\centerline{\hspace{-0.5in}
\epsfig{file=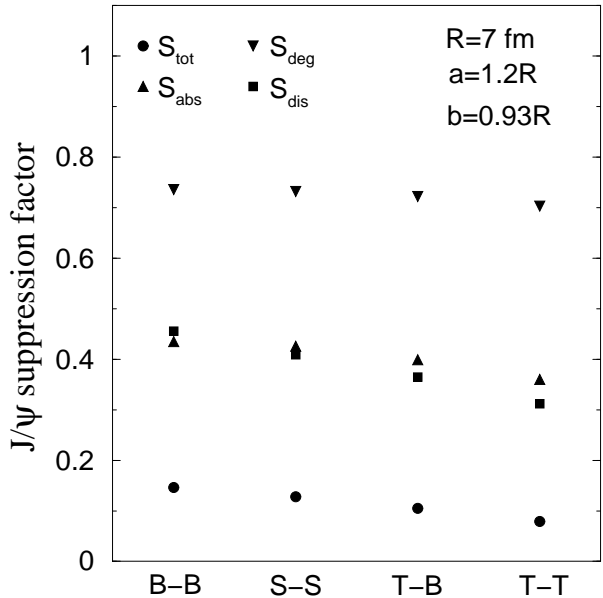,width=3.1in,height=3.1in,angle=-90}}
\vspace{0.2in}
\caption{Same as Fig. \ref{jp} but with an
effective color electric dissociation coefficient that depends on
the initial energy density. See the text for the details.}
\label{jpd}
\end{figure}

To include the effect of nuclear deformation on $J/\psi$
absorption due to the color electric field in the initial dense
matter, we need to take into account the dependence of the
dissociation coefficient $c_s$ on the collision geometry.  To
explore this dependence, we first show in Fig. \ref{energy} the
energy density determined from the JPCIAE model for a spherical
volume with a radius of 2 fm and located at the center of the
target nucleus as a function of time, which starts when the first
nucleon-nucleon collision occurs. The energy density reached in
the T-T collision is about 35\% higher than in the B-B collision.
We thus multiply the effective color electric
dissociation coefficient $c_s$ used previously for collisions of
spherical nuclei by the relative increase in the maximum energy
density, i.e., 1.2, 1.07, and 0.93 for the T-T, T-B, and B-B
collisions, respectively. Results from using the modified color
dissociation coefficients are shown in Fig. \ref{jpd}. We see that
this leads to a more pronounced dependence of the color
dissociation mechanism on the nuclear orientation than that of
both the nuclear absorption and energy degradation effects. As a
result, $J/\psi$ suppression in the T-T collision is now twice
stronger than in the B-B collision. This will be useful in
verifying the formation of a non-hadronic dense matter in the 
initial stage of relativistic nucleus-nucleus collisions once the 
orientation of the colliding nuclei is known.

In summary, a hadron and string cascade model, JPCIAE, has been
used to study the effect of nuclear deformation on $J/\psi$
suppression in U-U collisions at 200A GeV/c. A 35\% higher initial
energy density and a factor of two more $J/\psi$ suppression are
found in collisions if the major axes of both nuclei are along
the beam direction than if they are perpendicular to the beam
direction. Moreover, we have found a much more pronounced
deformation effect on $J/\psi$ suppression by the color electric
dissociation than by the hadronic absorption and the energy
degradation of leading nucleons. The study of $J/\psi$ suppression
in collisions of deformed nuclei will thus help find the 
signature for the formation of a quark-gluon plasma in relativistic 
nucleus-nucleus collisions.

We thank Che-Ming Ko, Bao-An Li and Bin Zhang for useful discussions 
and Torbj\"orn Sj\"ostrand for detailed instructions on using the PYTHIA 
programs. BHS would also like to thank Joe Natowitz for the
hospitality during his stay at the Cyclotron Institute of Texas
A\&M University. This work is supported in part by the National
Science Foundation Grant No. PHY-9870038, the Department of Energy
Grant DE-FG03-93-ER40773, the Robert A. Welch Foundation Grant No.
A-1358, and the Texas advanced Research Program Grant No.
FY97-010366-068. BHS is also partially supported by the Natural
Science Foundation of China and Nuclear Industry Foundation of
China.

\end{document}